\documentclass[12pt]{article}
\usepackage{amsfonts}
\usepackage{amsmath}
\usepackage{amsthm}
\usepackage{amssymb}
\usepackage{graphicx}
\usepackage{subfigure}
\usepackage{indentfirst}

\newcommand{\EE}{e^+e^-}

\newcommand{\taums}{m_{\tau}}

\newcommand{\beq}{\begin{equation}}
\newcommand{\eeq}{\end{equation}}
\newcommand\Fig[1]{Fig.~\ref{#1}}

\newcommand\Eq[1]{Eq.~\eref{#1}}

\def\eref#1{(\ref{#1})}

\begin{document}

\title{Discussion on upper limit of the precision for $\tau$ mass measurement\thanks{
Supported in part by National Natural Science Foundation of China (NSFC) under contracts Nos.: 11375206, 10775142, 10825524, 11125525, 11235011; the Ministry of Science and Technology
of China under Contract Nos.: 2015CB856700, 2015CB856705, and the CAS Center for Excellence in Particle Physics (CCEPP).} }

\author{X.H.Mo$^{1}$\footnote{E-mail:moxh@ihep.ac.cn}\\
  {\small 1 (Institute of High Energy Physics, Chinese Academy of
    Sciences, Beijing 100049, China )} }

\date{\today}
\maketitle

\begin{abstract}
$\tau$ lepton is one of three chareged leptons in nature, the measurements of its mass have been performed since its discovery. The present relative accuracy is already at the level of $10^{-4}$; more factors are still being studied in order to increase the accuracy. However, the available techniques for analysis and expectable luminosity from $\EE$ collider indicate that the precision upper limit of $\tau$ mass is almost reached, which means that brand new approaches should be looked for if the great improvement is yearned for.
\end{abstract}

\noindent
{\bf PACS} \hskip 0.25cm: 14.60.Fg, 13.35.Dx, 13.66.Jn \\

\section{Introduction}
\begin{figure}[htbp]
\begin{center}
\begin{minipage}{12cm}
\includegraphics[height=9cm,width=9.cm]{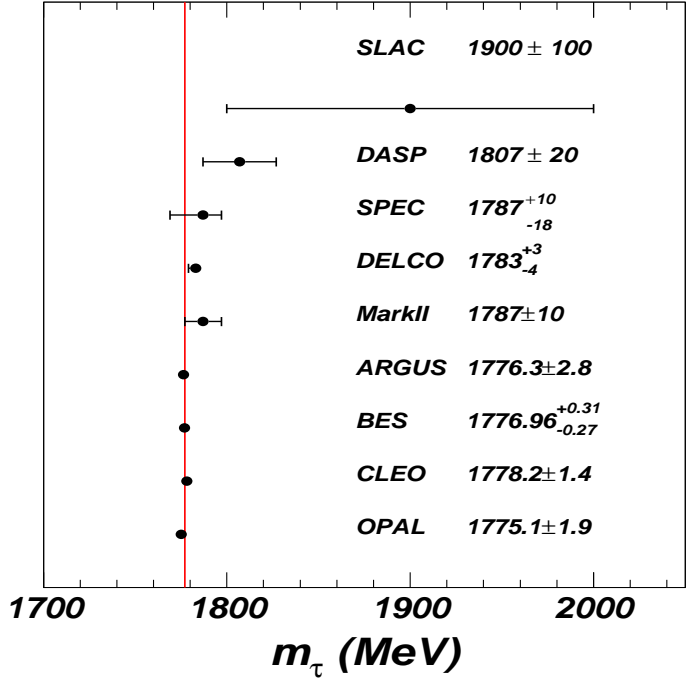}
\center (a) $\taums$ measured in the 20 century
\end{minipage}
\hskip 2cm
\begin{minipage}{12cm}
\includegraphics[height=7cm,width=9.cm]{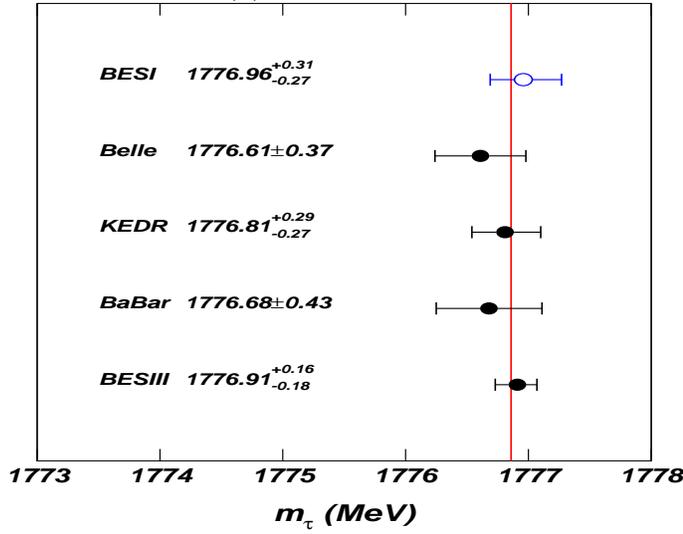}
\center (b) $\taums$ measured in the 21 century
\end{minipage}
\caption{\label{taumm}$\taums$ measured in the last and this century. In (a) the red line indicates the   average value of $\taums$ in PDG2000~\cite{PDG2000}: $m_{\tau} = 1777.03 ^{ + 0.30} _{-0.26}$ MeV. For comparison, the measured value from BES in 1996 is also plotted in (b) but with blank blue circle for distinction. The average value of $\taums$ in PDG2015~\cite{PDG2014}: $m_{\tau} = 1776.86 \pm 0.12$ MeV is also indicated by the red line in (b). It should be noted that since PDG1996~\cite{PDG1996,PDG1994}, the results from experiements performed before 1990 were removed except for the result of DELCO.}
\end{center}
\end{figure}

The history of $\tau$ mass ($\taums$) measurement has forty years. In the first experimental paper of $\tau$ lepton~\cite{Perl:1975bf}, $\taums$ is estimated to have a mass in the range from 1.6 to 2.0 GeV. Since then many experiments have been performed to measure $\taums$
\cite{Perl:1977se,Brandelik:1977xz,Bartel:1978ii,taudelco,Blocker1982,tauargus,taubes1,taubes2,taubes3,taucleo1,taucleo2,tauopal,taubelle,taukedr,taubabar,Ablikim:2014uzh}, the results are displayed in \Fig{taumm}.

The measurement results of $\taums$ in the 21 century are summarized in Table~\ref{tab:taums21cn}, where two results were acquired using the method of pseudo-mass while the others using the method of threshold scan. For the pseudo-mass method, the huge amount of data from B-factories are employed~\cite{taubelle}, good statistical accuracy is acquired, but large systematic uncertainty exists, which is mainly due to the absolute calibration of the particle momentum. For the threshold-scan method, the value of $\taums$ was extracted from the dependence of production cross section on beam energy. In the KEDR experiment~\cite{taukedr}, both resonant depolarization technique and Compton backscattering technique~\cite{Bogomyagkov} are used to determine the beam energy. All these techniques greatly decrease the uncertainty of beam energy.

\begin{table}[htb]
\caption{\label{tab:taums21cn}Measurement results of $\taums$ in the 21 century.}
\begin{tabular}{l|l|l|l|l} \hline \hline
Measured $\taums$                  & Year & Exp. Group & Data sample	& Method  \\
(value+statistic+systematic)       &      &            &             & \\  \hline
$1776.91\pm 0.12^{+0.10}_{-0.13}$  &2014  &BESIII~\cite{Ablikim:2014uzh} & 23.26 pb$^{-1}$ & Threshold-scan     \\
$1776.68\pm 0.12\pm 0.41 	   $   &2009  &Babar~\cite{taubabar}& 423 fb$^{-1}$ & Pseudo-mass \\
$1776.81^{+0.25}_{-0.23}\pm 0.15$  &2007  &KEDR~\cite{taukedr}  & 6.7 pb$^{-1}$ & Threshold-scan\\
$1776.61\pm 0.13\pm0.35	       $   &2007  &Belle~\cite{taubelle}& 414 fb$^{-1}$ & Pseudo-mass \\
 \hline \hline
\end{tabular}
\end{table}

Although the accuracy of results from above two methods are at the comparable level, it is obvious that the systematic errors already dominate for the pseudo-mass method, the enhancement of luminosity is already unappealing. For the threshold-scan method, both the statistic and systematic errors still seem to have room for the further improvement. Therefore, the subsequence discussions will be devoted to the threshold-scan method.

\section{Statistic aspect}\label{sxn:sttserr}
By the virtue of optimization theory~\cite{Mo:2015gla}, the relation between the absolute error of $\taums$ (denoted by $u$) and the total luminosity (denoted by $L$) is given by the following formula
\beq
u = \frac{A}{\sqrt{L}}~,
\label{eq:errlum}
\eeq
where $A$ is a constant depending on the cross section, the derivative of cross section to $\taums$, and some other quantities~\cite{Mo:2015gla}.

\begin{figure}[htbp]
\center
\includegraphics[height=8cm,width=8.cm]{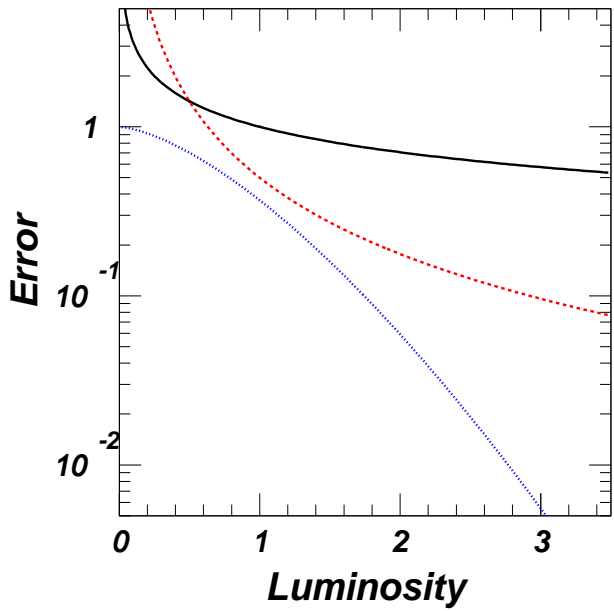}
\caption{\label{fig:lumerr}The relation between luminosity and error. The black solid line denotes the variation of $u$ against $L$; the red dash line the variation of $\nu$ against $L$; the blue dotted line the variation of $\xi$ against $L$. In calculation $A=1$ and $B=1$ are adopted. }
\end{figure}

An effective error variation that is called $L$-profit, is introduced and defined as
\beq
\nu \equiv \lim_{\Delta L \to 0} \left[ - \frac{\Delta u }{\Delta L}\right] = - \frac{du}{dL}
= \frac{A}{2} L^{-\frac{3}{2}}~.
\label{eq:efferr}
\eeq
The minus in the above definition indicates that when $L$ increases $u$ always decreases. This quantity discloses the variation of $u$ for a unit luminosity. Although it is reasonable to expect the reduction of error when the luminosity accumulates, but $L$-profit will also be small for larger $L$. Refer to Fig.~\ref{fig:lumerr}, when $u$ is considerably large, the enhancement of $L$ can effectively lead to the prominent improvement of $u$. But step by step, when $u$ is already accurate enough, a great mount of $L$ has to be consumed for reducing the error while the improvement is very limited, which is rather uneconomical.

The above analysis is not the whole story. As a matter of fact, the increment of $L$ means that more machine time must be used, more data must be dealt with, and more analysis must be performed. Moreover, the improvement of $u$ means that a lot of work concerning systematic uncertainty analysis must be finished, which in turn leads to the necessary demand for improvement of Monte Carlo simulation. A great amount of details has to be taken into account, sometime the upgrade of software is indispensable as well. Intuitively the work consumed for error improvement increases exponentially, and this is even more real when error is already small enough\footnote{There is no induction model or deduction proof for this exponential assumption, since the work done for a certain target is usually hard to be quantified. But for such an assumption, an interesting fact is worthy of mentioning which may be edificatory for understanding the exponential accretion. In the standard model the contribution to the anomalous magnetic moment $a_e$ comes from three types of interactions, electromagnetic, hadronic, and electroweak:
$$a_e=a_e(\mbox{QED})+a_e(\mbox{hadronic})+a_e(\mbox{elctroweak}).$$
The QED contribution can be evaluated by the perturbative expansion in $\alpha/\pi$:
$$a_e(\mbox{QED})=\sum_{n=1}^{\infty} \left( \frac{\alpha}{\pi} \right)^n a_e^{(2n)}~, $$
where $a_e^{(2n)}$ is finite due to the renormalizability of QED and may be written in general as
$$ a_e^{(2n)}= A_1^{(2n)}+ A_2^{(2n)} (m_e/m_{\mu})+  A_2^{(2n)} (m_{\mu}/m_{\tau})
+ A_3^{(2n)} (m_e/m_{\mu},m_{\mu}/m_{\tau} )$$
to show the mass-dependence explicitly. $A_1^{(2n)}$ can be calculated by QED theory perturbatively  with the increase of the order $n$. It is noticeable that the number of Feynman diagrams increase even more rapidly than the exponential increase with respect to $n$. The following table shows the interesting comparison~\cite{Aoyama:2012wj,bk:Greiner}.
\begin{center}
\begin{tabular}{llllll} \hline \hline
\multicolumn{6}{c}{Number of Feynman diagrams (N.F.D.) for $A_1^{(2n)}$} \\ \hline
 order      & $n=1$       & $n=2$      & $n=3$      &$n=4$      &$n=5$       \\
$A_1^{(2n)}$& $A_1^{(2)}$ & $A_1^{(4)}$& $A_1^{(6)}$&$A_1^{(8)}$& $A_1^{(10)}$ \\
N.F.D.      & 1           & 5          & 72         & 891       &12672       \\
$e^{2(n-1)}$& 1           & 7.39       & 54.60      & 403.43    &2980.96     \\ \hline \hline
\end{tabular}
\end{center}
}. Under such a consideration, a work-factor $D$ is designated as
\beq
D = e^{\frac{B}{u}} = e^{C\sqrt{L}}, \mbox{  where } C=\frac{B}{A}~.
\label{eq:labfactor}
\eeq
Here the constant $B$ will be related to concrete analysis procedure. Similar to \Eq{eq:efferr},  another effective error variation that is called $D$-profit, is introduced and defined as
\beq
\xi \equiv \lim_{\Delta D \to 0} \left[ - \frac{\Delta u }{\Delta D}\right] = -\frac{du/dL}{dD/dL}
= \frac{A^2}{B} ( L e^{C\sqrt{L}} )^{-1}~.
\label{eq:efferrb}
\eeq
If refer to Fig.~\ref{fig:lumerr}, it is fairly clear that the decrease of $\xi$ with $L$ is violent, which indicates the increase of luminosity will play smaller and smaller effect on the improvement of $\taums$. In another word the error improvement of $\taums$ in $\EE $ collider is no room using current method.

To have a concrete impression of above analysis, we compare two pedagogical experiments, the first one intends to decrease the error from 2 MeV to 1 MeV, while the second one intends to decrease the error from 0.2 MeV to 0.1 MeV. The coefficients $A$ and $B$ are chosen to be 1 for the unit of MeV. The ratio of aforementioned quantities is defined as
\beq
R^{i}_{X} \equiv \frac{X^i_a}{X^i_b}~,
\label{eq:defofr}
\eeq
where superscript $i$ indicates the first ($i=1$) or the second ($i=2$) experiment; subscript $a$ means the value after experiment, while $b$ the value before experiment. The symbol $X$ denotes the quantity $L$, $\nu$, $D$, and $\xi$ respectively. The evaluations are presented in Table~\ref{expexample}.

\begin{table}[htb]
\caption{\label{expexample}Evaluations for pedagogical experiments.} \center
\begin{tabular}{ccc} \hline \hline
Ratio value & First Exp.              & Second  Exp. \\
            & 2 $\to$ 1 MeV           & 0.2 $\to$ 0.1 MeV \\ \hline
$R_L=(u_b/u_a)^2$
            & $4   $                  & $ 4   $  \\
$R_{\nu}=(u_a/u_b)^3$
            & $1/8 $                  & $ 1/8 $  \\
$R_D=e^{(1/u_a-1/u_b)}$
            & $ 1.65$                 & $ 148 $  \\
$R_{\xi}=(u_a/u_b)^2\cdot e^{(1/u_b-1/u_a)}$
           & $ 0.15$                  & $ 1.69 \times 10^{-3}$  \\  \hline \hline
\end{tabular}
\end{table}

Based on Table~\ref{expexample} we know that if the error is improved to be one half of the previous experiment, four fold luminosity is required, and luminosity profit is one eighth. Actually according to quantities $L$ and $\nu$, as long as the improvement proportion is the same, $R_L$ and
$R_{\nu}$ are the same. This also implies that these two quantities can reflect neither the realistic working strength nor working profit. Turning to another two quantities $R_D$ and $R_{\xi}$, the situation becomes rather different. Although the improvement proportion is the same for the two experiments, the work have to be done for the second experiment is ninety times more than that for the first one, while $D$-profit for the second one is merely one ninetieth of the first one! This reminds us of the statement, that is, when $u$ is already accurate enough the further increase of luminosity is uneconomical and ineffective.

A remark is in order here. During the study of $\taums$ scan, BESIII collaboration have make methodical and meticulous studies on optimization process, to guarantee the effective usage of available luminosity~\cite{Mo:2015gla,wangyk2007,wangyk2009,wangbq2012,wangbq2013}. Nevertheless, this effort is not helped with the low $R_{\xi}$ issue.

\section{Limit due to systematic errors}\label{sxn:syserr}
For BESIII collaboration, before the actual scan of $m_{\tau}$, many studies have been made, one of which is the systematic estimation~\cite{wangyk2007} as listed in Table~\ref{sytuty}. It can be seen that the error due to energy scale dominates. To decrease such an uncertainty, starting from year 2007, a high accuracy beam energy measurement system (BEMS) located at the north crossing point (NCP) of BEPC-II was designed, constructed, and put into the commissioning at the end of 2010~\cite{bems2009,bems2010,bems,bems2}. The launching of the system is excellently well, two days are utilized to perform $\psi'$ resonance scan. The mass difference between the PDG2010 value~\cite{PDG2010} and the measured result by BEMS is $1 \pm 36$ keV, the deviation of which indicates that the relative accuracy of BEMS is at the level of 2 $\times 10^{-5}$~\cite{bems}.

\begin{table}[htb]
\caption{\label{sytuty}Systematic uncertainty estimation for $m_{\tau}$
measurement from Ref.~\cite{wangyk2007}.} \center
\begin{tabular}{ccc} \hline \hline
Source & $\delta m_\tau$           & $\delta m_\tau/m_\tau$  \\
       & ( $10^{-3}$ MeV)          &  ( $10^{-6}$ )         \\ \hline\hline
Luminosity            &   14.0     &    7.9   \\
Efficiency            &   14.0     &    7.9   \\
Branching Fraction    &   3.5      &    2.0   \\
Background            &   1.7      &    1.0   \\
Energy spread         &   3.0      &    1.7   \\
Theoretical accuracy  &   3.0      &    1.7   \\
Energy scale          & 100        &   56.3   \\ \hline Summation
 & 102 & 57.5  \\  \hline\hline
\end{tabular}
\end{table}

\begin{table}[tb]
\begin{center}
\caption{Summary of the  $\tau$ mass systematic errors. The meanings of each term can be found in Ref.~\cite{Ablikim:2014uzh}.}
\begin{tabular} {lc}
\hline \hline
Source & $\Delta m_{\tau}$ (MeV/$c^2$) \\
\hline
  Theoretical accuracy & 0.010 \\
  Energy scale & $^{+0.022}_{-0.086}$  \\
  Energy spread & 0.016 \\
  Luminosity & 0.006 \\
  Cut on number of good photons & 0.002 \\
  Cuts on PTEM and acoplanarity angle & 0.05 \\
  mis-ID efficiency    & 0.048 \\
  Background shape & 0.04 \\
  Fitted efficiency parameter & $^{+0.038}_{-0.034}$\\
  \hline
  Total & $^{+0.094}_{-0.124}$\\\hline\hline
\end{tabular}
\label{total-err}
\end{center}
\end{table}

In the actual data taking process, it is found that the measurement accuracy of BEMS is sensitive to the running status of accelerator~\cite{moxh2014}. At the end of 2011, a test scan of $\taums$ was performed. The integrated luminosity around of 23.26 pb$^{-1}$  is collected in the vicinity of $\tau$-pair threshold, together with 1.5 pb$^{-1}$ and 7.5 pb$^{-1}$ data sets for the $J/\psi$ and $\psi'$ resonance scans respectively, by virtue of which the systematic uncertainty due to beam energy is determined. The mass of $\tau$ lepton is measured from a maximum likelihood fit to the $\tau$-pair production cross-section to be $m_{\tau} =( 1776.91 \pm 0.12 ^{ + 0.10} _{- 0.13})$ MeV~\cite{Ablikim:2014uzh}, among which the systematic uncertainty due to energy scale also dominates all relevant terms, as shown in Table~\ref{total-err}. This tough situation is the same for all other $\EE$ colliders that adopt the threshold-scan method, which can not be circumvented either for the pseudo-mass method.

The crucial issue here lies in the uncertainty of energy scale. For BESIII at BEPCII, BEMS was established to control this uncertainty. The working scheme of this system can be described briefly as follows~\cite{principle}: firstly, the laser source provides the laser beam and the optics system focuses the laser beam and guides it to make head-on collisions with the electron (or positron) beam in the vacuum pipe, here the Compton backscattering process happens; after that the backscattering high energy photon will be detected by the HPGe detector, the detection capacity of which pins down the accuracy of beam energy. For the time being, the limit of relative accuracy for HPGe detector is at the level of $10^{-5}$, which in turn restricts the limit for determination of beam energy\footnote{For the present BEMS, the detector calibration is the main factor that restricts the detection accuracy. Generally speaking, there are two methods to calibrate the HPGe detector, one is the radiative source calibration. Several radiative sources are put nearby the detector and used to provide the simultaneous calibration lines during the data taking. The demerit of this method lies in that sometime not suitable sources can be found in the wide region of measurement. The other calibration method is to utilize the radiative sources from nuclear reactor, which can provide a great amount of calibration lines. But the shortcoming of this method consists in that calibration can be not done simultaneously with the data taking. However, in principle the calibration of detector can be improved further.}. Another juxtaposing restriction at the level of $10^{-5}$ comes from the long term stability of electric power supply of BEPCII~\cite{bk:bepcii}. This stability directly controls that of magnets, which  determines the energy of beam.

\section{Way out for accuracy improvement}\label{sxn:newway}
The preceding analyses obviously indicate that the efforts spent on $\EE$ collider are not to bring the promising improvement on the accuracy of $\taums$. Therefore, the new approach indeed needs to be figured out in order to increase the accuracy of $\taums$ significantly. Without any plausible precedent or theoretical implication, we would like to look at some relevant measurements with high accuracy to fish out some clues for $\taums$ experiments in the future.

\subsection{Experimental aspect}\label{sxn:expway}
According to PDG2012~\cite{PDG2012}
\beq
\begin{array}{rcl}
m_e      &=& 0.510998910 \pm 0.000000013~ \mbox{MeV}
 ~~~( \delta m_e/m_e \approx 2.554 \times 10^{-8} )~, \\
m_{\mu}  &=& 105.658367 \pm 0.000004~ \mbox{MeV}
 ~~~( \delta m_{\mu}/m_{\mu} \approx 3.786\times 10^{-8} )~,\\
\label{Eq.eumass}
\end{array}
\eeq

Such a highly precise determination of an electron's mass comes from measuring the ratio of the mass to that of a nucleus, so that the result is obtained in $m_u$ (atomic mass units). For example, by  comparing cyclotron frequencies of electrons and single $C^{6+}$ ions alternately confined to the same uniform magnetic field in a Penning trap~\cite{Farnham:1995zz}, the electron's mass is determined as
$$m_e = 0.000548 579911 1(12) m_u~~.$$
Then a convert factor $941.494013 \pm 0.000037 $ MeV/$m_u$ is used~\cite{PDG2012,Mohr:2000ie} to transform unit $m_u$ into MeV.

As far as the muon's mass is concerned, it is obtained from the muon-electron mass ratio as determined from the measurement of Zeeman transition frequencies in muonium ($\mu^+ e^-$ atom). Unfortunately, the life time of $\tau$ lepton is too short to form such an atom, so the mass of $\tau$ lepton can not be determined accurately in the similar way.

Anyway, it is fairly enlightened that we should find a relation of $\taums$ with another sensitive quantity which can be measured accurately, or their ratio can be measured accurately. This might be the new direction for the further accurate measurement of $\taums$.

\subsection{Theoretical aspect}\label{sxn:theway}
Similar to the situation of experiment, theoretically speaking, if a certain relation can be found for $\taums$, the excellent accuracy may be expected. Here we mention an interesting formula of masses of three leptons~\cite{Koide1,Koide2a,Koide2b,Koide2c}:
\beq
m_{e}+m_{\mu}+m_{\tau}=\frac{2}{3} (\sqrt{m_{e}}+\sqrt{m_{\mu}}+\sqrt{m_{\tau}})^2~,
\label{Eq.koidefml}
\eeq
In the light of which it is easy to find
\beq
\sqrt{m_{\tau}}=2(\sqrt{m_{e}}+\sqrt{m_{\mu}})+\sqrt{3(m_{e}+m_{\mu})+12\sqrt{m_{e} \cdot m_{\mu}}}~.
\label{Eq.koidefm2}
\eeq
Using this formula together with the mass values for electron and muon in \Eq{Eq.eumass}
the $\taums$ can be obtained together with the corresponding error as follows
\beq
m_{\tau}= 1776.968884 \pm 0.000058~ \mbox{MeV}
~~~( \delta m_{\tau}/m_{\tau} \approx 3.26\times 10^{-8} ).
\label{Eq.exptaums}
\eeq

Unfortunately, the formula~\eref{Eq.koidefml} itself is not obtained from the first principle of particle physics, so this kind of efforts can only be treated as an attempt direction in the future\footnote{
Among the mysteries of particle physics, one of the most elusive is the mass spectrum that can't be predicted from the elementary principle, which is also the most unsatisfactory feature of the present Standard Model. The arbitrariness of mass assignment triggers the decades of studies~\cite{Nambu1952,Barut1980,Bettini2009,Boya2011}, among which several interesting results emerge. A very simple formula (i.e. \Eq{Eq.koidefml}) between the masses of the three charged leptons was stumbled over by Yoshio Koide at the beginning of 1980's when he was working on some composite models of quarks and leptons~\cite{Koide1,Koide2a,Koide2b,Koide2c}. Further relevant studies are performed~\cite{Rivero:2005vj,Koide:2015ura,Barut1979,Gsponera1996,Rosen:1995kt}, although the impact is small, some interesting results are worth of mention, one of which is Barut's formula for leptons~\cite{Barut1979}, viz.
$$ m(N) = m_e \left(1 +\frac{3}{2}\alpha^{-1} \sum_{n=0}^{N} n^4 \right)~, $$
where $m_e$ is the mass of electron and $\alpha \approx 1/137 $ the electromagnetic fine structure constant. The masses of electron, muon, and tauon correspond to $N =0$, 1, and 2, respectively.

Besides the dynamic-inspired heuristic researches, there are also phenomenological effects towards a systematic comprehension of the variety of elementary particles, two of which are to be recapitulated herein. The first one is the $\alpha$-quantized scheme suggested by M.H.Mac Gregor~\cite{ Gregor:2005, Gregor:1974, Gregor:1976, Gregor:1990}. Based on  investigating 36 metastable threshold-state (lifetimes great than $10^{-21}$ second), $\alpha$-spaced groups are found, which exhibit a factor-of-3 c-quark-to-b-quark ``flavor structure'' and a pervasive factor-of-2 ``hyperfine structure''. And these structures can be parameterized by two-variable-quantum scheme.

The second one is the shell-quantized scheme proposed by P.~Palazzi~\cite{Palazzi:2003,Palazzi:2004,Palazzi:2005}. Edifying by nuclear shell-structure, the stability analysis of the mass spectrum reveals the clear mesonic shell-structure. A type-specific mass unit in the vicinity of 35 MeV is verified for all the meson-states listed by the PDG with fairly good accuracy except for only 5 abnormal mesons. Furthermore, the quantization of mass unit can be geometrically qualified by the face-centered-cubic model of the nucleus~\cite{Cook:1987zz}.

Up till now, no breakthrough achievement has been acquired. Nevertheless, the explorations related to the mass spectrum indicates another direction on the understanding of high energy physics, which might bring an unprecedented result some day in the future.
}.

\section{Summary}\label{sct:sumary}

In this paper, the issue concerned the upper limit of the error of $\taums$ measurement is discussed from unconventional point of view. The main conclusions acquired are summarized below:

\begin{enumerate}
\item Statistically speaking, the increase of luminosity have no effective effect on the accuracy improvement of $\taums$. Moreover, from point view of efficiency and profitability, the input-output ratio will considerably decrease as the accuracy of $\taums$ increases.
\item Systematically speaking, the limit at the level of $10^{-5}$ of both calibration of HPGe detector and energy stability of accelerator circumscribes the accuracy limit of $\taums$. For experiments at $\EE$ collider, this is the insurmountable limit for the systematic uncertainty.
\item The absolutely distinctive train of thought from both theoretical and experimental aspects is needed to seek a new direction to increase the accuracy of $\taums$.
\end{enumerate}

\end{document}